\documentclass[twocolumn,showpacs,preprintnumbers,amsmath,amssymb]{revtex4}
\usepackage{epsfig}

\usepackage{graphicx}
\usepackage{dcolumn}
\usepackage{bm}

\begin{document}
\newcommand{\be}{\begin{equation}}
\newcommand{\ee}{\end{equation}}
\newcommand{\bea}{\begin{eqnarray}}
\newcommand{\eea}{\end{eqnarray}}
\newcommand{\disp}{\displaystyle}
\preprint{APS/123-QED}

\title{Two Qubit Entanglement in $XYZ$ Magnetic Chain with DM Antisymmetric Anisotropic Exchange Interaction}

\author{Zeynep Nilhan GURKAN}
\email{nilhangurkan@iyte.edu.tr}
\author{Oktay K. PASHAEV}%
 \email{oktaypashaev@iyte.edu.tr}
\affiliation{%
Department of Mathematics, Izmir Institute of Technology,
Urla-Izmir, 35430, Turkey
}%

\date{\today}

\begin{abstract}
In the present paper we study two qubit entanglement in the most
general  $XYZ$ Heisenberg magnetic chain with (non)homogeneous
magnetic fields and the  DM anisotropic antisymmetric exchange
interaction, arising from the spin-orbit coupling . The model
includes all known results as particular cases, for both
antiferromagnetic and ferromagnetic  $XX, XY, XXX, XXZ, XYZ$
chains. The concurrence of two qubit thermal entanglement and its
dependence on anisotropic parameters, external magnetic field and
temperature are studied in details. We found that in all cases,
inclusion of the DM interaction, which is responsible for weak
ferromagnetism in mainly antiferromagnetic crystals and spin
arrangement in low symmetry magnets, creates (when it does not
exist) or strengthens (when it exists) entanglement in $XYZ$ spin
chain. This implies existence of a relation between arrangement of
spins and entanglement, in which the DM coupling plays an
essential role. It  suggests also that anisotropic antisymmetric
exchange interaction could  be  an efficient control parameter of
entanglement in the general $XYZ$ case.
\end{abstract}

\pacs{Valid PACS appear here}
\maketitle
\newtheorem{thm}{Theorem}[subsection]
\newtheorem{cor}[thm]{Corollary}
\newtheorem{lem}[thm]{Lemma}
\newtheorem{prop}[thm]{Proposition}
\newtheorem{defn}[thm]{Definition}
\newtheorem{rem}[thm]{Remark}
\newtheorem{prf}[thm]{Proof}
\section{\label{sec:level1}Introduction }

Entanglement property has been discussed at the early years of
quantum mechanics as specifically quantum mechanical nonlocal
correlation \cite{Schrodinger}- \cite{Bell} and it becomes
recently a key point of quantum information theory \cite{Bennet}.
For entangled subsystems, the whole state vector cannot be
separated into a product of the states of the subsystems, and the
last ones are no longer independent even if they are far spatially
separated. A measurement on one subsystem not only gives
information about the other subsystem, but also provides
possibilities of manipulating it. Therefore in quantum
computations the entanglement becomes main tool of information
processing, such as quantum cryptography, teleportation and etc.

For realizing quantum logic gates, several models have been
proposed and demonstrated by experiments in cavity QED, ion trap,
and NMR \cite{zheng}, \cite{atac}. Due to intrinsic pairwise
character of the entanglement, in all these cases important is to
find entangled qubit pairs. Generic two qubit state is
characterized by 6 real degrees of freedom. While separable two
qubit state has only four degrees of freedom. It is clear that
single qubit gates are unable to generate entanglement in an $N$
qubit system, because starting from separable state we will obtain
another separable state with transformed by gates separable
qubits. Then to prepare an entangled state one needs inter qubit
interactions which is a two qubit gate. The well known example of
two qubit gate generating entanglement is Controlled Not (CNOT)
gate \cite{Benenti}. Moreover realization of two qubit controlled
gates is a necessary requirement for implementation of the
universal quantum computation. For this purpose we also need
interacting qubits. A simple example of two qubit interaction is
described by the Ising interaction $J \sigma_1^z \sigma_2^z $
between spin $1/2$ particles. More general interaction between two
qubits is given by the Heisenberg magnetic spin chain model. This
model have been extensively studied during several decades,
experimentally in condensed matter systems \cite{wigen} and
theoretically as exactly solvable many body problems (Bethe,
Baxter and others) \cite{lieb}, \cite{baxter}. Now they become
promising to realize quantum computation and information
processing by generating entangled qubits and constructing quantum
gates. Quantum spin chains where proposed as medium through which
quantum information could propogate as a pulse \cite{linden}.
Recently several proposals discussing quantum entanglement of two
qubits in such models has been considered \cite{nielsen}. It was
noticed that in isotropic Heisenberg spin chain $XXX$ model spin
states are unentangled in the ferromagnetic case ($J<0$), while
for the antiferromagnetic case ($J>0$) entanglement occurs for
sufficiently small temperature $T< T_c=\frac{2J}{k\ln 3}$.
Important point is how to increase entanglement in situation where
it exists already or to create entanglement in situation when it
does not exist. Certainly this can be expected from a
generalization of bilinear spin-spin interaction form. Around 50
years ago to explain weak ferromagnetism of antiferromagnetic
crystals ($\alpha- Fe_2 O_3, MnCO_3 $ and $CrF_3$), which has been
controversial problem for a decade, Dzialoshinski
\cite{Dzialoshinski} from phenomenological arguments and Moriya
\cite{Moriya} from microscopic grounds have introduced anisotropic
antisymmetric exchange interaction, the Dzialoshinski-Moriya (DM)
interaction, expressed by
$$\vec D \cdot [\vec S_1 \times \vec S_2].$$
This interaction arises from extending the Anderson`s theory of
superexchange interaction by including  the spin orbit coupling
effect \cite{Moriya} , and it is important not only for weak
ferromagnetism but also for the spin arrangement in
antiferromagnets of low symmetry. In the present paper we show
that the Dzialoshinski-Moriya interaction plays an essential role
for entanglement of two qubits in magnetic spin chain model of
most general $XYZ$ form. We find that in all cases, inclusion of
the DM interaction creates (when it does not exist) or strengthens
(when it exists) entanglement. In particular case of isotropic
Heisenberg $XXX$ model discussed above, inclusion of this term
increases entanglement for antiferromagnetic case and even in
ferromagnetic case,  for sufficiently strong coupling $D> (kT
sinh^{-1}e^{|J|/kT}-J^2)^{1/2}$, it creates entanglement.  These
results imply existence of an intimate relation between weak
ferromagnetism of mainly antiferromagnetic crystals and the spin
arrangement in antiferromagnets of low symmetry, with entanglement
of spins. Moreover it shows that the DM interaction could be an
efficient control parameter of entanglement in the general $XYZ$
model.
\section{$XYZ$ Heisenberg Model}
The Hamiltonian of $XYZ$ model for $N$ qubits is  \bea H =
\sum_{i=1}^{N-1}\frac{1}{2}[J_x \,\, \sigma^x_i \sigma^x_{i+1} +
J_y \,\, \sigma^y_i \sigma^y_{i+1} + J_z \,\, \sigma^z_i
\sigma^z_{i+1} +  \\ (B+b)\, \sigma^z_i + (B-b) \, \sigma^z_{i+1}
+ \vec D\cdot (\vec{\sigma_i} \times \vec{\sigma}_{i+1})]
\nonumber \eea where $B, b$- external homogeneous and
nonhomogeneous magnetic fields respectively, the last term is the
DM coupling. Choosing $ \frac{\vec{D}}{2}= \frac{D}{2} \cdot \vec
z $ the Hamiltonian for two qubits becomes \bea H &=& \frac{1}{2}
[J_x \, \sigma^x_1 \sigma^x_2 + J_y \,\, \sigma^y_1 \sigma^y_2 +
J_z \,\, \sigma^z_1 \sigma^z_2 + (B+b) \, \sigma^z_1
\nonumber\\&+&(B-b)\, \sigma^z_2 + D (\sigma^x_1 \sigma^y_2 -
\sigma^y_1 \sigma^x_2 ) ]\label{Hamiltonian1} \eea
  and in the matrix form {\small{\be H=
\left[
         \begin{array}{cccc}
           \disp \frac{J_z}{2}+ B & 0 & 0 & \disp \frac{J_x - J_y}{2} \\
           0 & \disp -\frac{J_z}{2}+ b &\disp \frac{J_x + J_y}{2}+ iD & 0 \\
           0 &\disp \frac{J_x + J_y}{2}- iD & \disp -\frac{J_z}{2}- b & 0 \\
          \disp \frac{J_x - J_y}{2} & 0 & 0 & \disp \frac{J_z}{2}- B \\
         \end{array}
       \right].
\nonumber \ee }} To study thermal entanglement firstly we need to
obtain all the eigenvalues and eigenstates of the Hamiltonian
(\ref{Hamiltonian1}): $ H|\Psi_i \rangle = E_i | \Psi_i \rangle ,
\,\,\,\, (i= 1,2,3,4). \label{Hamiltonianeqn}$  The eigenvalues
(energy levels) are: \bea E_1&=& \frac{J_z}{2}- \mu
\,\,\,\,\,\,\,\,\,\,\, E_3= -\frac{J_z}{2}- \nu \nonumber \\E_2&=&
\frac{J_z}{2}+ \mu
 \,\,\,\,\,\,\,\,\,\,\,
E_4= -\frac{J_z}{2}+ \nu\nonumber \eea  where $\frac{J_x -
J_y}{2}\equiv J_-$, $\frac{J_x + J_y}{2}\equiv J_+$, $\mu \equiv
\sqrt{B^2 + J_-^2}$, $\nu \equiv \sqrt{b^2+ J_+^2+D^2}$ and
corresponding wave functions are {\small \begin{eqnarray}
|\Psi_1\rangle &=& \frac{1}{\sqrt{2(\mu^2+ B\nu)}} \left[
     \begin{array}{c}
           J_- \\
           0 \\
           0 \\
           -(B+\mu) \\
         \end{array}
       \right] \nonumber \\
  |\Psi_2\rangle &=&\frac{1}{\sqrt{2(\mu^2-B \nu)}} \left[
     \begin{array}{c}
           J_- \\
           0 \\
           0 \\
           -(B-\mu) \\
         \end{array}
       \right] \nonumber \\
|\Psi_3\rangle &=&\frac{-i}{\sqrt{2(\nu^2+ b \nu)}} \left[
    \begin{array}{c}
           0\\
           J_++iD \\
            -(b+\nu)\\
          0 \\
         \end{array}
        \right]\nonumber \\
|\Psi_4\rangle &=&\frac{1}{\sqrt{2(\nu^2-b\nu)}} \left[
       \begin{array}{c}
           0\\
           J_++iD \\
            -(b-\nu)\\
          0 \\
         \end{array}
      \right]\nonumber \end{eqnarray} }

For $B=0,\, b=0 ,\, D=0$ the wave functions reduce to the Bell
states {\small \bea |\Psi_2\rangle \longrightarrow |B_0\rangle &=&
\frac{1}{\sqrt{2}}\,(|00 \rangle +|11 \rangle)
          \\
|\Psi_4\rangle \longrightarrow |B_1\rangle &=&
\frac{1}{\sqrt{2}}\,(|01 \rangle +|10 \rangle)\\ |\Psi_3\rangle
\longrightarrow |B_2\rangle &=& \frac{1}{\sqrt{2}}\,(|01 \rangle
-|10 \rangle) \\
|\Psi_1\rangle \longrightarrow |B_3\rangle &=&
\frac{1}{\sqrt{2}}\,(|00 \rangle -|11 \rangle) \eea }
  The
state of the system at thermal equilibrium is determined by the
 density matrix
 \be \rho(T)= \frac{e^{-H/ k T}}{Tr[e^{-H /
k T}]}=\frac{e^{-H/ k T}}{Z} , \label{densitymatrix}\ee where $Z$
is the partition function, $k$ is Boltzmann's constant and $T$ is
the temperature. Then for Hamiltonian (\ref{Hamiltonian1})
 we find
 {\small \bea e^{-H / k T}&=& I +
\left(\frac{-H}{kT}\right) + \frac{1}{2!}
\left(\frac{-H}{kT}\right)^2+ ... +  \frac{1}{n!}
\left(\frac{-H}{kT}\right)^n+ ...\nonumber\\
&=& \left[
         \begin{array}{cccc}
            A_{11} & 0 & 0 & A_{14} \\
           0 &   A_{22} & A_{23} & 0 \\
           0 &  A_{32} &  A_{33} & 0 \\
           A_{41} & 0 & 0 &  A_{44}  \\
         \end{array}
       \right]
  \eea}
  where
\small{ \bea A_{11}&=& -e^{\frac{- J_z}{2kT}}\left[ \cosh
\frac{\mu}{kT}-\frac{B}{\mu}\sinh \frac{\mu}{kT}\right] \nonumber\\
A_{14}&=& - e^{\frac{- J_z}{2kT}}\frac{J_-}{\mu} \sinh
\frac{\mu}{kT}\nonumber\\
A_{22}&=& e^{\frac{J_z}{2kT}}\left[ \cosh
\frac{\nu}{kT}-\frac{b}{\nu}\sinh \frac{\nu}{kT}\right]\nonumber\\
A_{23}&=& -e^{\frac{J_z}{2kT}}\frac{J_++iD}{\nu}\sinh
\frac{\nu}{kT}
\nonumber\\
A_{32}&=& -e^{\frac{J_z}{2kT}}\frac{J_+-iD}{\nu}\sinh
\frac{\nu}{kT}
 \nonumber\\
A_{33}&=& e^{\frac{J_z}{2kT}}\left[ \cosh
\frac{\nu}{kT}+\frac{b}{\nu}\sinh \frac{\nu}{kT}\right] \nonumber\\
 A_{41}&=&-e^{-\frac{J_z}{2kT}}\frac{J_-}{\mu} \sinh
\frac{\mu}{kT}\nonumber\\
A_{44} &=& e^{-\frac{J_z}{2kT}}\left[ \cosh
\frac{\mu}{kT}+\frac{B}{\mu}\sinh \frac{\mu}{kT}\right] \eea and
\be Z=Tr[e^{-H/kT}]= 2 \left[ e^{\frac{-J_z}{2kT}} \cosh
\frac{\mu}{kT}+ e^{\frac{J_z}{2kT}} \cosh \frac{\nu}{kT}
\right].\nonumber\ee}  As $\rho(T)$ represents a thermal state,
the entanglement in this state is called the \textbf{\emph{thermal
entanglement}}. The concurrence $C$ (the order parameter of
entanglement) is defined as \cite{Wooters1},  \cite{Wooters2} \be
C= max \{\lambda_1-\lambda_2-\lambda_3-\lambda_4, 0\} \ee where
$\lambda_i \, (i=1,2,3,4)$ are the ordered square roots of the
eigenvalues of the operator \be \rho_{12}= \rho (\sigma^y \otimes
\sigma^y) \rho^{*}(\sigma^y \otimes \sigma^y) \ee and
$\lambda_1>\lambda_2>\lambda_3>\lambda_4 > 0$. The concurrence is
bounded function $0 \le C \le 1$.When the concurrence $C=0$,
states are unentangled; when $C=1$, states are maximally
entangled. In our case:

{\small \begin{eqnarray} \lambda_{1,2} &=&
\frac{e^{\frac{-J_z}{2kT}}}{Z} \left|\sqrt{1+
\frac{J_-^2}{\mu^2}\sinh^2 \frac{\mu}{kT}}\mp
\frac{J_-}{\mu} \sinh \frac{\mu}{kT}\right|\\
\lambda_{3,4}&=& \frac{ e^{\frac{J_z}{kT}}}{Z} \left|\sqrt{1+
\frac{J_+^2+D^2}{\nu^2}\sinh^2 \frac{\nu}{kT}}\mp
\frac{\sqrt{J_+^2+D^2}}{\nu} \sinh \frac{\nu}{kT}\right| \nonumber
\end{eqnarray}}
where $\mu \equiv \sqrt{B^2 + J_-^2}$, $\nu \equiv \sqrt{b^2+
J_+^2+D^2}$. Before calculating the concurrence for the general
$XYZ$ case (Section 7) it is instructive to consider particular
reductions of $XYZ$ model and compare corresponding concurrences
with known results.
\section{Ising Model}
Let $J_x=J_y=0$ and $J_z \neq 0$ and the Hamiltonian is \be H=
\frac{1}{2}[ J_z\,\, \sigma^z_1 \sigma^z_2 +(B+b)\, \sigma^z_1 +
(B-b)\, \sigma^z_2 + D(\sigma^x_1 \sigma^y_2- \sigma^y_1
\sigma^x_2 )] \nonumber\ee
  The eigenvalues are {\small \begin{eqnarray}  \lambda_{1,2}&=&
\frac{e^{\frac{-J_z}{2kT}}}{Z}\\
\lambda_{3,4} &=& \frac{ e^{ \frac{J_z}{2kT}}}{Z}\left|\sqrt{1 +
\frac{D^2}{\nu^2} \sinh^2 \frac{\nu}{kT}} \mp \frac{D}{\nu}\sinh
\frac{\nu}{kT}\right|\nonumber
 \end{eqnarray}}
where $\mu=B$, $\nu=\sqrt{b^2+D^2}$ and {\small \be
Z=Tr[e^{-H/kT}]= 2 \left[ e^{\frac{-J_z}{2kT}} \cosh \frac{B}{kT}+
e^{\frac{J_z}{2kT}} \cosh \frac{\sqrt{b^2+D^2}}{kT} \right].
\nonumber \ee }
\subsection{\label{sec:level2}Pure Ising Model ($B=0, b=0, D=0 $)}
\subsubsection{\label{sec:level3}Antiferromagnetic Case ($J_z>0$):}
The ordered eigenvalues are \be \lambda_1=\lambda_2
\frac{e^{J_z/2kT}}{Z}> \lambda_3=\lambda_4=\frac{e^{-J_z/2kT}}{Z}.
\ee where $Z= 4\cosh\frac{ J_z}{2kT}$ and the concurrence is
     \be C_{12}= max \{ \frac{- e^{J_z/2kT}}{ 2\cosh \frac{J_z}{2kT}}, 0
     \}=0 \ee and there is no entanglement.
     \subsubsection{Ferromagnetic Case ($J_z<0$):}
     The ordered eigenvalues are \be \lambda_1=\lambda_2
\frac{e^{|J_z|/2kT}}{Z}>
\lambda_3=\lambda_4=\frac{e^{-|J_z|/2kT}}{Z}. \ee where $Z=
4\cosh\frac{ |J_z|}{2kT}$ and the concurrence is
    \be  C_{12}= max \{ \frac{- e^{-|J_z|/2kT}}{ 2\cosh \frac{|J_z|}{2kT}}, 0
\}=0 \ee It means that in both antiferromagnetic and ferromagnetic
cases there is no entanglement in pure Ising Model for any $T$.
\subsection{Ising Model with Homogeneous Magnetic Field ($B
\neq 0, b=0, D=0 $)}
 The ordered eigenvalues are \be \lambda_1=\lambda_2
\frac{e^{J_z/2kT}}{Z} ,\,\,\,\,
\lambda_3=\lambda_4=\frac{e^{-J_z/2kT}}{Z}. \ee where $Z=
2\left[e^{-J_z/2kT}\cosh\frac{B}{kT}+ e^{J_z/2 kT}\right]$ and the
concurrence
 \be C_{12}= max \{ \frac{- e^{J_z/2kT}}{
2(e^{-J_z/2kT}\cosh\frac{B}{kT}+ e^{J_z/ kT})}, 0 \}=0 \ee and
there is no entanglement.
\subsection{Ising Model with Nonhomogeneous Magnetic Field ($B= 0,
b \neq 0, D=0 $)} The concurrence is \be \disp C_{12}= max \{
\frac{- e^{J_z/2kT}}{ 2(e^{-J_z/2kT}+ e^{J_z/
kT}\cosh\frac{b}{kT})}, 0 \}=0 \ee and there is no entanglement.

As we can see in pure Ising model and including homogeneous (B)
and nonhomogeneous (b) magnetic fields no entanglement occurs
\cite{vedral}, \cite{terzis}, \cite{childs}.
\subsection{Ising Model with DM Coupling ($B=0, b=0, D\neq 0 $)}

The eigenvalues are \bea \lambda_1&=& \frac{e^{(J_z+2D)/2kT}}{Z}
,\,\,\,\,\, \lambda_2 = \frac{e^{(J_z-2D)/2kT}}{Z} \\
\lambda_3&=&\lambda_4=\frac{e^{-J_z/2kT}}{Z}. \eea where $Z=
2\left[e^{J_z/2kT}\cosh\frac{D}{kT}+ e^{-J_z/ 2kT}\right]$.

\subsubsection{Antiferromagnetic Case ($J_z>0$):}
Ordering the eigenvalues $\lambda_1> \lambda_2 > \lambda_3=
\lambda_4$ we have the concurrence
$$ C_{12} = max \{\frac{\sinh
\frac{|D|}{kT}-e^{-J_z/kT}}{\cosh\frac{|D|}{kT}+ e^{-J_z/ kT}}, 0
\}
 $$ Then $C_{12}=0$ (no entanglement) if $\sinh \frac{|D|}{kT}\leq e^{-J_z/
 kT}$.
 When $\sinh \frac{|D|}{kT}> e^{-J_z/
 kT}$ the states are entangled  \be C_{12}
= \frac{\sinh \frac{|D|}{kT}-e^{-J_z/kT}}{\cosh\frac{|D|}{kT}+
e^{-J_z/ kT}}. \label{91}\ee Moreover  states become more
entangled for low temperatures: maximally entangled for any $D$
and $T=0$ so that ($ \lim_{kT \rightarrow 0}C_{12}=1 $)
  and for stronger DM coupling $ \lim_{D \rightarrow \infty}
C_{12}=1 .$  With $T$ growing, $D_{min}= kT \sinh^{-1} e^{-J_z/
kT} $ is growing so that we need to increase $D$ to have entangled
states.
 \subsubsection{Ferromagnetic Case
($J_z<0$):}
\begin{description}
 \item [a)] With weak DM coupling $|D|<|J_z|$ there is no entanglement
      Ordering the eigenvalues
        $\lambda_3=\lambda_4>\lambda_1 > \lambda_2$ we have the concurrence  {\small \be C_{12}=  max \{\frac{-\cosh
\frac{|D|}{kT}}{\cosh\frac{|D|}{kT}e^{-|J_z|/kT}+ e^{|J_z|/ kT}}, 0
\}=0 \ee }
    \item [b)] With strong DM coupling $|D|>|J_z|$
    Ordering the eigenvalues
         $\lambda_1> \lambda_3=\lambda_4> \lambda_2$ and the
        we have the concurrence
      {\small   \begin{eqnarray} C_{12}= max \{\frac{\sinh
\frac{|D|}{kT}-e^{|J_z|/2kT}}{\cosh\frac{|D|}{kT}+ e^{|J_z|/
2kT}}, 0 \}
\end{eqnarray}}
\end{description}
Then $C_{12}=0$ (no entanglement) if $\sinh \frac{|D|}{kT}\leq
e^{|J_z|/
 kT}$. When $\sinh \frac{|D|}{kT}> e^{|J_z|/
 kT}$ or $|D|> |J_z|+ \frac{kT}{2}\ln (1+ e^{{-2|J_z|}/{kT}})$ the states are entangled {\small \be C_{12}
= \frac{\sinh \frac{|D|}{kT}-e^{|J_z|/kT}}{\cosh\frac{|D|}{kT}+
e^{|J_z|/ kT}}. \label{96}\ee } Moreover  states become more
entangled for low temperatures
 $ \lim_{kT \rightarrow 0}C_{12}=1 $
  and for stronger DM coupling $ \lim_{D \rightarrow \infty}
C_{12}=1 .$ As we can see there is entanglement even in
ferromagnetic case with sufficiently strong DM coupling.
Comparison of (\ref{91}) and (\ref{96}) shows that in
anti-ferromagnetic case, states can be more easily entangled then
in the ferromagnetic one
\section{$XX$ Heisenberg Model}
 For $J_z=0, J_x=J_y\equiv J$,
 the Hamiltonian is \be H= \frac{1}{2}[ J \,
(\sigma^x_1  \sigma^x_2 +  \sigma^y_1 \sigma^y_2) + (B+b)\,
\sigma^z_1 + (B-b)\,  \sigma^z_2 + D(\sigma^x_1  \sigma^y_2-
\sigma^y_1 \sigma^x_2 )] \nonumber\ee The eigenvalues are {\small
\begin{eqnarray}
\lambda_{1,2}&=& \frac{1}{Z} \\
\lambda_{3,4} &=& \frac{1}{Z} \left|\sqrt{1+
\frac{J^2+D^2}{\nu^2}\sinh^2 \frac{\nu}{kT}}\mp
\frac{\sqrt{J^2+D^2}}{\nu} \sinh \frac{\nu}{kT}\right| \nonumber
\end{eqnarray} } where $\nu=\sqrt{J^2+b^2+D^2}$ and $$
Z=Tr[e^{-H/kT}]= 2 \left[\cosh \frac{B}{kT}+  \cosh \frac{\nu}{kT}
\right]. $$
\subsection{Pure $XX$ Heisenberg Model ($B=0, b=0, D=0 $)}
The eigenvalues are
 \be \lambda_1=
\frac{e^{J/kT}}{Z} ,\,\,\,\, \lambda_2=\lambda_3=
\frac{1}{Z},\,\,\,\, \lambda_4= \frac{e^{-J/kT}}{Z}. \ee
\subsubsection{Antiferromagnetic Case $J>0$}
The ordered eigenvalues are $\lambda_1
> \lambda_2=\lambda_3> \lambda_4$ and the concurrence is $\disp C_{12}= max \{ \frac{\sinh
\frac{J}{kT}-1}{\cosh \frac{J}{kT}+1}, 0 \} $ and for
\begin{description}
\item [a)] $\sinh \frac{J}{kT}> 1$ \, $\Rightarrow$ $\disp
C_{12}=\frac{\sinh \frac{J}{kT}-1}{\cosh \frac{J}{kT}+1}$ so that
\be  \lim_{T \rightarrow 0} C_{12}=1 \ee \item [b)] $\sinh
\frac{J}{kT}\leq 1$ $\Rightarrow$ $C_{12}=0 $ there is no
entanglement for \be T> \underbrace{\frac{J}{k} [\sinh^{-1}1]^{-1}
}_{T_C}\ee
\end{description}
\subsubsection{Ferromagnetic Case $J<0$}
he eigenvalues are \be \lambda_1= \frac{e^{-|J|/kT}}{Z} ,\,\,\,\,
\lambda_2=\lambda_3= \frac{1}{Z},\,\,\,\, \lambda_4=
\frac{e^{|J|/kT}}{Z}. \ee $\lambda_4
> \lambda_2= \lambda_3> \lambda_1$
and the concurrence is \be C_{12} = max \{ \frac{\sinh
\frac{|J|}{kT}-1}{\cosh \frac{|J|}{kT}+1}, 0 \} \ee and
\begin{description}
\item [a)] $\sinh \frac{|J|}{kT}> 1$ \, $\Rightarrow$ $\disp
C_{12}=\frac{\sinh \frac{|J|}{kT}-1}{\cosh \frac{|J|}{kT}+1}$
\item [b)] $\sinh \frac{|J|}{kT} \leq 1$ \,$\Rightarrow$ $C_{12}=0
$ no entanglement for \be \disp T> \underbrace{\frac{|J|}{k}
[\sinh^{-1}1]^{-1} }_{T_c}\ee
\end{description}
In both cases states are entangled at sufficiently small
temperature $T< T_C = \frac{|J|}{K}[sinh^{-1}1]^{-1}$.

\subsection{$XX$ Heisenberg Model with Magnetic Field ($B \neq 0, b=0, D=0 $ )}
The eigenvalues are \be \lambda_1= \frac{e^{J/kT}}{Z} ,\,\,\,\,
\lambda_2=\lambda_3= \frac{1}{Z},\,\,\,\, \lambda_4=
\frac{e^{-J/kT}}{Z}. \ee where \be Z= 2 \left[\cosh \frac{B}{kT}+
\cosh \frac{J}{kT} \right].\ee and the concurrence is $\disp
C_{12}= max \{ \frac{\sinh \frac{J}{kT}-1}{\cosh \frac{J}{kT}+
\cosh \frac{B}{kT}}, 0 \}$ and
\begin{description}
\item [a)] $\sinh \frac{|J|}{kT}> 1$ $\Rightarrow$ $\disp C_{12}=
\frac{\sinh \frac{J}{kT}-1}{\cosh \frac{J}{kT}+ \cosh
\frac{B}{kT}}
 $ \item [b)]$\sinh \frac{|J|}{kT} \leq 1, \,\,\,
C_{12}=0 $ no entanglement for \be T> \underbrace{\frac{|J|}{K}
[\sinh^{-1}1]^{-1} }_{T_C} \ee
\end{description}
It shows that inclusion of magnetic field does not change the
critical temperature for concurrence in both anti-ferromagnetic
and ferromagnetic cases. Pairwise entanglement in $N$- qubit $XX$
chain and experimental realization of $XX$ model has been
discussed in \cite{zheng}, \cite{atac}, \cite{wang3}, \cite{xi1}
and \cite{xi2}.
\subsection{$XX$ Heisenberg Model with DM Coupling ($B=b=0,\,\,D \neq
0$)}
  The eigenvalues are \be \lambda_{1,2}=
\frac{1}{Z}, \,\,\,\,\, \lambda_3= \frac{e^{-\beta/kT}}{Z}, \,\,\,\,
\lambda_4=\frac{e^{\beta/kT}}{Z} \ee where $\beta>0,\,\,\, \beta=
\sqrt{J^2+ D^2}$ and $Z= 2(1+ \cosh \frac{\beta}{kT})$. The ordered
eigenvalues are $\lambda_4>\lambda_3>\lambda_1= \lambda_2$ and the
concurrence is {\small \begin{eqnarray} C_{12} &=& max
\{\frac{\sinh\frac{\nu}{kT}-1}{\cosh \frac{\nu}{kT}+1},0\}
\end{eqnarray}}
where $\nu = \sqrt{J^2+D^2}:$
\begin{description}
 \item [a)] $\sinh\frac{\nu}{kT}>1$
$\Rightarrow$ $\disp C_{12}=\frac{\sinh\frac{\nu}{kT}-1}{\cosh
\frac{\nu}{kT}+1} $
    \item [b)] $\sinh\frac{\nu}{kT}\leq 1$  $\Rightarrow$    $C_{12}=0$ there is no
    entanglement.
\end{description}
Entanglement increases with growth of DM coupling in both
anti-ferromagnetic and ferromagnetic cases.

\section{$XY$ Heisenberg Model} For $J_z=0, J_x \neq J_y$, the Hamiltonian is \be H=
\frac{1}{2}[ J_x \, \sigma^x_1  \sigma^x_2 +J_y \sigma^y_1
\sigma^y_2 + (B+b)\, \sigma^z_1 + (B-b)\, \sigma^z_2 +
D(\sigma^x_1 \sigma^y_2- \sigma^y_1 \sigma^x_2 )]\nonumber \ee The
eigenvalues are {\small
\begin{eqnarray} \lambda_{1,2}
&=& \frac{1}{Z} \left|\sqrt{1+ \frac{J_-^2}{\mu^2}\sinh^2
\frac{\mu}{kT}}\mp
\frac{J_-}{\mu} \sinh \frac{\mu}{kT}\right|\\
\lambda_{3,4} &=& \frac{1}{Z} \left|\sqrt{1+
\frac{J_+^2+D^2}{\nu^2}\sinh^2 \frac{\nu}{kT}}\mp
\frac{\sqrt{J_+^2+D^2}}{\nu} \sinh \frac{\nu}{kT}\right| \nonumber
\end{eqnarray}} where $\mu= \sqrt {B^2+J_-^2}$, $\nu=
\sqrt{b^2+D^2+J_+^2}$, $J_{\pm}\equiv \frac{J_x+J_y}{2}$ and
{\small \be Z=Tr[e^{-H/kT}]= 2 \left[\cosh \frac{\mu}{kT}+ \cosh
\frac{\nu}{kT} \right].\ee }
\subsection{Pure $XY$ Heisenberg Model ($B=0, b=0, D=0 $)}
The eigenvalues are \be \lambda_1= \frac{e^{J_- / kT}}{Z},\,\,\,
\lambda_2= \frac{e^{-J_- / kT}}{Z},\,\,\, \lambda_3= \frac{e^{J_+
/ kT}}{Z},\,\,\, \lambda_4= \frac{e^{-J_+ / kT}}{Z},\,\,\,
\nonumber\ee where \be Z= 2 \left[\cosh \frac{J_- }{kT} + \cosh
\frac{J_+}{kT}\right].\ee

For $J_x=J(1+\gamma)$ and $J_y= J(1-\gamma)$ so that $J_+= J,
J_-=J\gamma $,  the eigenvalues are \be \lambda_1= \frac{e^{J
\gamma / kT}}{Z},\,\,\, \lambda_2= \frac{e^{-J\gamma /
kT}}{Z},\,\,\, \lambda_3= \frac{e^{J / kT}}{Z},\,\,\, \lambda_4=
\frac{e^{-J / kT}}{Z},\,\,\,  \ee
\subsubsection{Anti-ferromagnetic Case $J_x>0$ and
$J_y>0$ } The ordered eigenvalues are
        $\lambda_3>\lambda_1> \lambda_2 > \lambda_4$ and the concurrence \be C_{12} =max \{ \frac{\sinh
\frac{J_+}{kT} - \cosh \frac{J_-}{kT}}{\cosh \frac{J_-}{kT}+ \cosh
\frac{J_+}{kT}}, 0 \} \label{concXYpure}\ee
\begin{description}
    \item [a)] $\sinh \frac{J_+}{kT} > \cosh \frac{J_-}{kT}$
    $\Rightarrow$
$\disp C_{12}= \frac{\sinh \frac{J_+}{kT} - \cosh
\frac{J_-}{kT}}{\cosh \frac{J_-}{kT}+ \cosh \frac{J_+}{kT}}
$,\,\,\, ($ \lim_{T \rightarrow 0} C_{12}=1 $)
    \item [b)]$\sinh \frac{J_+}{kT}\leq \cosh \frac{J_-}{kT}$ $\Rightarrow$
    $C_{12}=0$ there is no entanglement.
\end{description}
In Fig. 1, we plot the concurrence $C_{12}$ in $XY$ Heisenberg
antiferromagnet as function of $\frac{J_+}{kT}$ and
$\frac{J_-}{J_+}$

\begin{figure}[h]
\begin{center}
\epsfig{figure=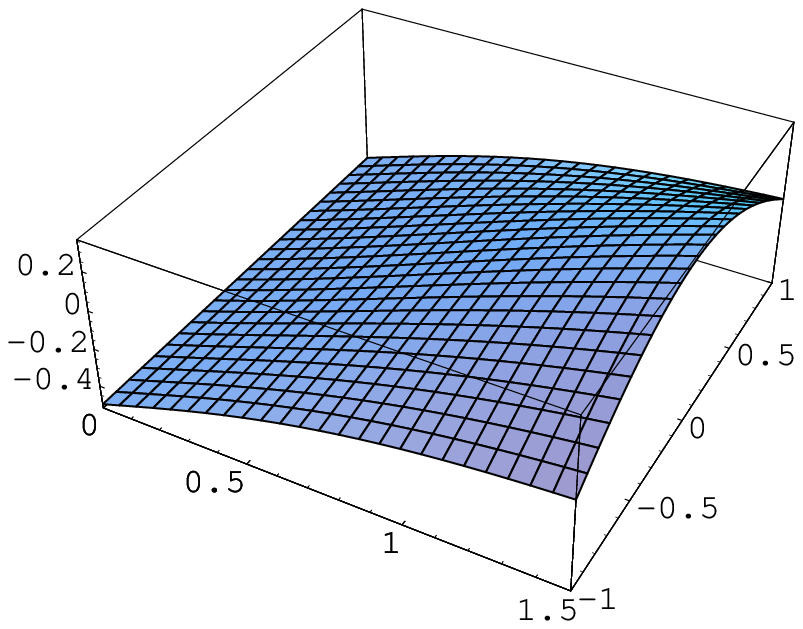,height=4cm,width=4cm}
\end{center}
\caption{Concurrence $C_{12}$ in $XY$ antiferromagnet as function
of $\frac{J_+}{kT}$ and $\frac{J_-}{J_+}$} \label{Bound4}
\end{figure}

\subsubsection{ Ferromagnetic Case $J_x<0$ and $J_y<0$
}{\small \be
 C_{12}=  max \{ \frac{\sinh \frac{|J_-|}{kT} - \cosh \frac{J_+}{kT}}{\cosh
\frac{|J_-|}{kT}+ \cosh \frac{J_+}{kT}},0
\}\label{concXYpureferro}\ee }
\begin{description}
    \item [a)] $\sinh \frac{|J_-|}{kT} > \cosh \frac{J_+}{kT}$
{\small \be C_{12}= \frac{\sinh \frac{J_+}{kT} - \cosh
\frac{|J_-|}{kT}}{\cosh \frac{|J_-|}{kT}+ \cosh \frac{J_+}{kT}}
\ee \be \lim_{T \rightarrow 0} C_{12}=1 \ee }
    \item [b)] $\sinh \frac{|J_-|}{kT}\leq \cosh \frac{J_+}{kT}$
    $\Rightarrow$
    $C_{12}=0$ there is no entanglement.
\end{description}
In Fig. 2, we plot the concurrence $C_{12}$ in $XY$ Heisenberg
ferromagnet as function of $\frac{J_+}{kT}$ and
$\frac{|J_-|}{|J_+|}$.

\begin{figure}[h]
\begin{center}
\epsfig{figure=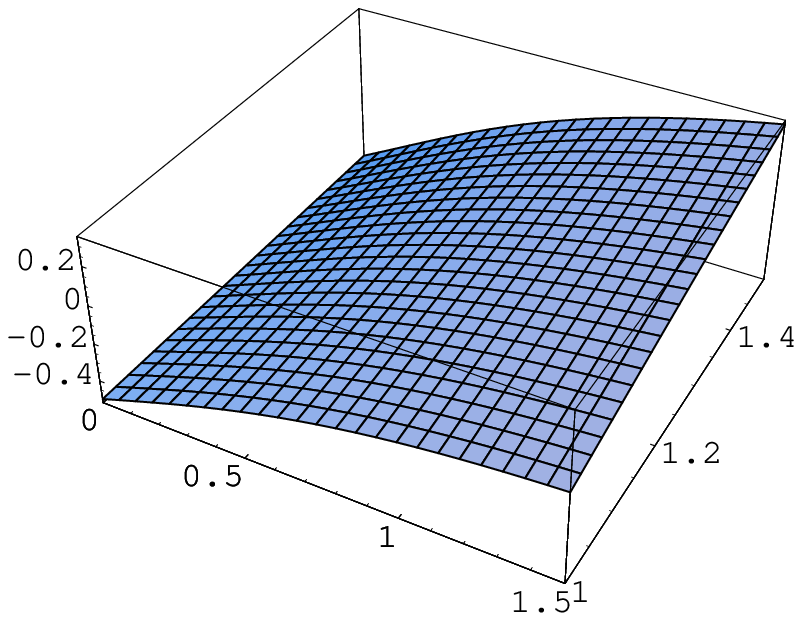,height=4cm,width=4cm}
\end{center}
\caption{Concurrence $C_{12}$ in XY ferromagnet as function of
$\frac{J_+}{kT}$ and $|\frac{J_-}{J_+}|$}
\end{figure}
Thermal entanglement in $XY$ chain was studied in \cite{wang1},
\cite{hamieh} and \cite{kamta} in the presence of external
magnetic field $B$ and in \cite{sun} by introducing non-uniform
magnetic field $b$.

\subsection{$XY$ Heisenberg Model with DM Coupling ($B = 0, b = 0 , D
\neq 0$)} The eigenvalues are \bea \lambda_1&=&
\frac{e^{J_-/kT}}{Z},\,\,\, \lambda_2= \frac{e^{-J_-/kT}}{Z}
\\ \lambda_3&=& \frac{e^{\sqrt{J_+^2+D^2}/kT}}{Z},\,\,\, \lambda_4=
\frac{e^{-\sqrt{J_+^2+D^2}/kT}}{Z}\eea where \be Z= 2\left[\cosh
\frac{|J_-|}{kT}+ \cosh \frac{\sqrt{J_+^2+ D^2}}{kT}\right]\ee
\subsubsection{Antiferromagnetic Case} The concurrence is {\small
\be C_{12}=  max \{ \frac{\sinh \frac{\sqrt{J_+^2+D^2}}{kT} -
\cosh \frac{J_-}{kT}}{\cosh \frac{\sqrt{J_+^2+D^2}}{kT}+ \cosh
\frac{J_-}{kT}},0 \}.\ee } It shows that for any temperature $T$
we can adjust sufficiently strong DM coupling $D$ to have
entanglement.
\begin{description}
    \item [a)] $\sinh \frac{\sqrt{J_+^2+D^2}}{kT} > \cosh
    \frac{J_-}{kT}$ $\Rightarrow$ \be C_{12}=  \frac{\sinh \frac{\sqrt{J_+^2+D^2}}{kT} - \cosh
\frac{J_-}{kT}}{\cosh \frac{\sqrt{J_+^2+D^2}}{kT}+ \cosh
\frac{J_-}{kT}}\ee
    \item [b)] $\sinh
\frac{\sqrt{J_+^2+D^2}}{kT} \leq \cosh
    \frac{J_-}{kT}$ $\Rightarrow$
     $C_{12}=0$ there is no entanglement.
\end{description}
\subsubsection{Ferromagnetic Case}
Ferromagnetic case gives the same result as anti-ferromagnetic
case. Comparison with pure XY model (\ref{concXYpure}) and
(\ref{concXYpureferro}) shows that level of entanglement is
increasing with growing DM coupling D and $C_{12}=1$ when $D
\rightarrow \infty$.
\section{$XXX$ Heisenberg Model} For $J_x=J_y=J_z \equiv J$, the
Hamiltonian is \bea H&=& \frac{1}{2}[ J (\sigma^x_1 \sigma^x_2 +
\sigma^y_1 \sigma^y_2 +  \sigma^z_1 \sigma^z_2) +(B+b)\,
\sigma^z_1
\\&+& (B-b)\,  \sigma^z_2 + D(\sigma^x_1 \sigma^y_2- \sigma^y_1
\sigma^x_2 )]. \nonumber\eea The eigenvalues are {\small
\begin{eqnarray}
\lambda_{1,2}&=& \frac{e^{-J/2kT}}{Z}\\
\lambda_{3,4} &=& \frac{ e^{J/2kT}}{Z} \left|\sqrt{1+
\frac{J^2+D^2}{\nu^2}\sinh^2 \frac{\nu}{kT}}\mp
\frac{\sqrt{J^2+D^2}}{\nu} \sinh \frac{\nu}{kT}\right|\nonumber
\end{eqnarray}}
where \be Z= 2\left[e^{-J/2kt} \cosh \frac{B}{kT}+ e^{J/2kT} \cosh
\frac{\sqrt{J^2+b^2+D^2}}{kT}\right] \ee

 \subsection{Pure XXX Model ($B=0, b=0,
D=0$)} The eigenvalues are
 \be \lambda_{1,2}= \frac{e^{-J/2kT}}{Z},\,\,\, \lambda_3=
\frac{e^{-J/2kT}}{Z},\,\,\,  \lambda_4= \frac{e^{3J/2kT}}{Z}\ee
where  \be Z= 2\left[e^{-J/2kT}+ e^{J/2kT}\cosh
\frac{J}{kT}\right] \ee

\subsubsection{Antiferromagnetic case ($J>0$):} The
concurrence is  \be C_{12}= max \{\frac{
e^{2J/kT}-3}{e^{2J/kT}+3},0 \}\label{XXX}\ee
\begin{description}
    \item [a)] $\sinh \frac{J}{kT} > e^{-J/kT}$
{\small \be C_{12}= \frac{\sinh \frac{J}{kT} -
e^{-J/kT}}{e^{-J/2kT}+ \cosh \frac{J}{2kT}}  \label{152}\ee } For
sufficiently small temperature $T< \frac{2J}{k\ln 3}$ entanglement
occurs and $ \lim_{T \rightarrow 0} C_{12}=1$
    \item [b)]$\sinh \frac{J}{kT} \leq e^{-J/kT}$ $\Rightarrow$
     $C_{12}=0$ there is no entanglement.
\end{description}
\subsubsection{Ferromagnetic case ($J<0$):} The
concurrence $\disp C_{12}= max \{\frac{-\cosh
\frac{|J|}{kT}}{\cosh \frac{|J|}{kT}+ e^{|J|/kT}},\,0 \}=0$ and no
entanglement occurs.

Thus, for ferromagnets, spins are always disentangled, while
entanglement is observed for antiferromagnets \cite{arnesen},
\cite{nielsen}.

\subsection{$XXX$ Heisenberg Model with Magnetic Field ($B \neq
0$)} The eigenvalues are
 \be \lambda_{1,2}= \frac{e^{-J/2kT}}{Z},\,\,\, \lambda_3=
\frac{e^{-J/2kT}}{Z},\,\,\,  \lambda_4= \frac{e^{3J/2kT}}{Z}\ee
\be Z= 2\left[e^{-J/2kT}\cosh \frac{B}{kT}+ e^{J/2kT}\cosh
\frac{J}{kT}\right]\ee
\subsubsection{Antiferromagnetic case ($J>0$):} The
concurrence \be C_{12}=  max \{\frac{e^{2J/kT}-3}{e^{2J/kT}+1+
2\cosh \frac{B}{kT}},0 \}\label{XXXB}\ee
\begin{description}
    \item [a)]$\sinh \frac{J}{kT} > e^{-J/kT}$,\,\, $\disp C_{12}= \frac{\sinh \frac{J}{kT} - e^{-J/kT}}{e^{-J/2kT} \cosh
\frac{B}{kT}+ \cosh \frac{J}{2kT}}\label{158} $ For sufficiently
small temperature $T< \frac{2J}{k\ln 3}$ entanglement occurs and
$\lim_{T \rightarrow 0} C_{12}=1$. Comparison of (\ref{XXXB}) with
(\ref{XXX}) shows that inclusion of the magnetic field $B$ does
not change the critical value but decreases the level of
entanglement.
    \item [b)] $\sinh \frac{J}{kT} < e^{-J/kT}$
     $\Rightarrow$ $C_{12}=0$ there is no entanglement.
\end{description}
\subsubsection{Ferromagnetic case ($J<0$):} $\disp
C_{12}= max \{\frac{-\cosh \frac{|J|}{kT}}{\cosh \frac{|J|}{kT}+
e^{|J|/kT}\cosh \frac{B}{kT}},\,0 \}=0$ and no entanglement
occurs. Therefore inclusion of magnetic field does not change the
result. Entanglement in $XXX$ Heisenberg model with magnetic field
has been studied in \cite{arnesen}.

\subsection{$XXX$ Heisenberg Model with DM Coupling ($B = 0, b
= 0, D\neq 0$)} The eigenvalues are \bea \lambda_{1,2}&=&
\frac{e^{-J/2kT}}{Z},\,\,\,\,\ \lambda_3= \frac{e^{(J-
2\sqrt{J^2+D^2})/2kT}}{Z}\\ \lambda_4&=& \frac{e^{(J+
2\sqrt{J^2+D^2})/2kT}}{Z}\eea where \be Z= 2\left[e^{-J/2kT}+
e^{J/2kT}\cosh \frac{\sqrt{J^2+D^2}}{kT}\right] \ee

\subsubsection{Antiferromagnetic Case ($J>0$):}
The concurrence is {\small \be C_{12}= max \{ \frac{\sinh
\frac{\sqrt{J^2+D^2}}{kT}- e^{-J/kT}}{e^{-J/kT}+ \cosh
\frac{\sqrt{J^2+D^2}}{kT}}, 0 \} \ee }
\begin{description}
    \item [a)] $ \sinh \frac{\sqrt{J^2+D^2}}{kT} >
e^{-J/kT}$ \be C_{12}= \frac{\sinh \frac{\sqrt{J^2+D^2}}{kT}-
e^{-J/kT}}{e^{-J/kT}+ \cosh \frac{\sqrt{J^2+D^2}}{kT}} \ee For a
given temperature, when \be D> \sqrt{kT sinh^{-1}e^{-J/kT}-J^2}\ee
there is entanglement.
    \item [b)] $ \sinh \frac{\sqrt{J^2+D^2}}{kT}\leq
e^{-J/kT}$ $\Rightarrow$ $C_{12}=0$ there is no entanglement.
\end{description}
\subsubsection{Ferromagnetic Case ($J<0$):} The
concurrence is {\small \be C_{12}= max \{ \frac{\sinh
\frac{\sqrt{J^2+D^2}}{kT}- e^{|J|/kT}}{e^{|J|/kT}+ \cosh
\frac{\sqrt{J^2+D^2}}{kT}}, 0 \} \ee }
\begin{description}
    \item [a)] $ \sinh \frac{\sqrt{J^2+D^2}}{kT} >
e^{|J|/kT}$ {\small \be C_{12}= \frac{\sinh
\frac{\sqrt{J^2+D^2}}{kT}- e^{|J|/kT}}{e^{|J|/kT}+ \cosh
\frac{\sqrt{J^2+D^2}}{kT}} \ee } For a given temperature, when \be
D> \sqrt{kT sinh^{-1}e^{|J|/kT}-J^2} \ee there is entanglement.
    \item [b)] $ \sinh \frac{\sqrt{J^2+D^2}}{kT}<
e^{|J|/kT}$  $\Rightarrow$ $C_{12}=0$
\end{description}

As we can see inclusion of DM coupling $D$ in XXX case  increases
entanglement in antiferromagnetic case and even create
entanglement in ferromagnetic case. Thermal enatnglement and
entanglement teleportation in $XXX$ Heisenberg chain with DM
interaction has been studied in \cite{zheng}.

\section{$XXZ$ Heisenberg Model} For $J_x=J_y=J \neq J_z$ the Hamiltonian is \bea
H&=& \frac{1}{2}[ J (\sigma^x_1 \sigma^x_2 + \sigma^y_1 \sigma^y_2
+ \Delta \, \sigma^z_1 \sigma^z_2) +(B+b)\, \sigma^z_1 \\&+&
(B-b)\, \sigma^z_2 + D(\sigma^x_1 \sigma^y_2- \sigma^y_1
\sigma^x_2 )]. \nonumber\eea where $\Delta \equiv {J_z}/{J}$. The
eigenvalues are {\small
\begin{eqnarray}
\lambda_{1,2}&=& \frac{e^{\frac{-J_z}{2kT}}}{Z}\\
 \lambda_{3,4}
&=& \frac{ e^{\frac{J_z}{2kT}}}{Z} \left|\sqrt{1+
\frac{J^2+D^2}{\nu^2}\sinh^2 \frac{\nu}{kT}}\mp
\frac{\sqrt{J^2+D^2}}{\nu} \sinh \frac{\nu}{kT}\right|
\nonumber\end{eqnarray}} where $\mu=B$, $\nu= \sqrt{J^2+D^2+b^2}$
and $$ Z= 2\left[e^{-J_z/2kT}\cosh \frac{B}{kT}+ e^{J_z/2kT} \cosh
\frac{\nu}{kT}\right]$$. \subsection{Pure $XXZ$ Heisenberg Model
($B=0, b=0, D=0$)} In this case the eigenvalues become \be
\lambda_{1,2}= \frac{e^{-J_z/2kT}}{Z},\,\,\, \lambda_3=
\frac{e^{(J_z-2J)/2kT}}{Z},\,\,\,  \lambda_4=
\frac{e^{(J_z+2J)/2kT}}{Z}\ee where $\beta=J$ and $Z=
2\left[e^{-J_z/2kT}+ e^{J_z/2kT}\cosh \frac{J}{kT}\right]$.
\subsubsection{Antiferromagnetic Case ($J>0$):}
  For  $|\Delta|<1$ weak anisotropy ($\Delta>0$ easy axis, $\Delta<0$ easy plane) and  $\Delta>1$ strong
 anisotropy, the concurrence is
{\small \be C_{12}= max \{ \frac{\sinh \frac{J}{kT}- e^{-J_z/kT}}{
\cosh \frac{J}{kT}+ e^{-J_z/kT}}, 0 \} \ee}
\begin{description}
    \item [a)] $\sinh \frac{J}{kT}>
 e^{-J_z/kT}$ {\small \be C_{12}= \frac{\sinh \frac{J}{kT}- e^{-J_z/kT}}{
\cosh \frac{J}{kT}+ e^{-J_z/kT}} \label{concpureXXZ}\ee}
    \item [b)] $\sinh \frac{J}{kT}\leq
e^{-J_z/kT}$ $\Rightarrow$ $C_{12}=0$,  no entanglement.

From above formulas follow that for sufficiently small temperature
$T$ the states are entangled.

For $\Delta \leq -1$ the concurrence is {\small \be C_{12}= max \{
\frac{-\cosh \frac{|J|}{kT}}{ \cosh \frac{|J|}{kT}+ e^{|J_z|/kT}},
0 \}=0 \ee } and no entanglement.

For $\Delta= 1$ the anisotropic model reduces the
 isotropic  $XXX$ model, and the concurrence reduces to
\be
 C_{12}= max \{ \frac{e^{2J/kT}-3}{e^{2J/kT}+3}, 0 \}=0 \ee
 When the temperature is larger than the critical temperature $T_C=
 \frac{2J}{ k \ln 3} $ the thermal entanglement disappears.
\end{description}
\subsubsection{Ferromagnetic Case ($J<0$):}
 For  $\Delta < 1$ The concurrence is \be C_{12}= max \{ \frac{\sinh
\frac{|J|}{kT}- e^{|J_z|/kT}}{ \cosh \frac{|J|}{kT}+
e^{|J_z|/kT}}, 0 \} \ee
\begin{description}
    \item [a)] $\sinh \frac{|J|}{kT}>
e^{|J_z|/kT}$ \be C_{12}= \frac{\sinh \frac{|J|}{kT}-
e^{|J_z|/kT}}{ \cosh \frac{J}{kT}+ e^{|J_z|/kT}}\ee
    \item [b)]$\sinh \frac{|J|}{kT}\leq
e^{|J_z|/kT}$ $\Rightarrow$ $C_{12}=0$ \end{description}

 For  $\Delta \geq 1$
the concurrence is \be C_{12}= max \{ \frac{-\cosh
\frac{|J|}{kT}}{ \cosh \frac{|J|}{kT}+ e^{|J_z|/kT}}, 0 \}=0 \ee
and no entanglement.
\subsection{$XXZ$ Heisenberg Model with DM Coupling ($B =0, b = 0, D\neq 0$)}
The eigenvalues are \bea \lambda_{1,2}&=&
\frac{e^{-J_z/2kT}}{Z},\,\,\, \lambda_3= \frac{e^{(J_z-2
\sqrt{J^2+D^2})/2kT}}{Z}\\ \lambda_4&=& \frac{e^{(J_z+2
\sqrt{J^2+D^2})/2kT}}{Z}\eea where  \be Z= 2\left[e^{-J_z/2kT}+
e^{J_z/2kT} \cosh \frac{\sqrt{J^2+D^2}}{kT}\right] \ee
\subsubsection{Antiferromagnetic case ($J>0$):} The concurrence is
{\small \be C_{12}= max \{ \frac{\sinh \frac{\sqrt{J^2+D^2}}{kT}-
e^{-J_z/kT}}{ \cosh \frac{\sqrt{J^2+D^2}}{kT}+e^{-J_z/kT} }, 0 \}
\ee }
\begin{description}
    \item [a)] $\sinh
\frac{\sqrt{J^2+D^2}}{kT}> e^{-J_z/kT}$ {\small \be C_{12}=
\frac{\sinh \frac{\sqrt{J^2+D^2}}{kT}- e^{-J_z/kT}}{ \cosh
\frac{\sqrt{J^2+D^2}}{kT}+e^{-J_z/kT} } \ee }
    \item [b)] $\sinh
\frac{\sqrt{J^2+D^2}}{kT}< e^{-J_z/kT}$ $\Rightarrow$  $C_{12}=0$
Comparison with (\ref{concpureXXZ}) shows that with growth of $D$
entanglement increases.
\end{description}
\subsubsection{Ferromagnetic Case ($J<0$):}
\begin{description}
\item [a)] For small $D<D_c= \sqrt{J_z^2-J^2}$ no entanglement.
\item [b)] For $D>D_c$ the concurrence is {\small \be C_{12}= max
\{ \frac{\sinh \frac{\sqrt{J^2+D^2}}{kT}- e^{|J_z|/kT}}{ \cosh
\frac{\sqrt{J^2+D^2}}{kT}+e^{|J_z|/kT} }, 0 \} \ee } and
entanglement increases with growing $D$.
\end{description}
Entanglement for $XXZ$ Heisenberg model was considered in
\cite{qiang} and effect of $DM$ interaction on $XXZ$ model in
\cite{wang2}.

\section{$XYZ$ Heisenberg Model}  \subsection{Pure XYZ Model ($B=0,
b=0, D=0$)} The eigenvalues are \bea \lambda_1&=& \frac{ e^{(-J_z-
2J_-)/2kT}}{Z},\,\, \lambda_2= \frac{ e^{(-J_z+ 2J_-)/2kT}}{Z}\\
\lambda_3&=&\frac{ e^{(J_z- 2J_+)/2kT}}{Z} ,\,\, \lambda_4=\frac{
e^{(J_z+ 2J_+)/2kT}}{Z}\eea where \be Z= 2\left[e^{-J_z/2kT} \cosh
\frac{J_-}{kT}+ e^{J_z/2kT} \cosh \frac{J_+}{kT}\right]
\ee\subsubsection{Antiferromagnetic Case :} $J_z>J_y>J_x>0
\Rightarrow $ $J_+>0, J_= -|J_-|<0$. The biggest eigenvalue is
$\lambda_4= \frac{e^{\frac{|J_z|+2|J_+|}{2kT}}}{Z}$ and the
concurrence is {\small
$$C_{12}=max \{\frac{\sinh \frac{J_+}{kT}-\cosh
\frac{J_-}{kT}e^{-J_z/kT} }{\cosh \frac{J_+}{kT}+ \cosh
\frac{J_-}{kT}e^{-J_z/kT} },0\}.  $$ } Then entanglement occurs
when {\small \be f(T)=\sinh \frac{J_+}{kT}-\cosh
\frac{J_-}{kT}e^{-J_z/kT}>0. \label{inequa}\ee}  This formula
shows that entanglement increases with lowering temperature.
\begin{figure}[h]
\begin{center}
\epsfig{figure=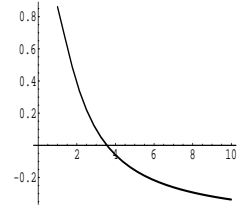,height=3cm,width=3cm}
\end{center}
\caption{Concurrence in $XYZ$ antiferromagnet as function of T}
\end{figure}In Fig. 3,
we plot function $f(T)$ for $(J_z,J_y,J_x)=(3,2,1)$. It shows
entanglement for $T<T_c$. In addition, from (\ref{inequa}) we have
entanglement increasing with growing anisotropy $J_+$ and
decreasing with growing anisotropy $J_-$. Moreover it increases
with growing $J_z$.
\subsubsection{Ferromagnetic Case :} Let $J_z<J_y<J_x<0$ then
$J_+=-|J_+|$, $J_-=|J_-|<0$ and $J_z=-|J_z|$. The biggest
eigenvalue is {\small $$\lambda_1=\frac{
e^{(|J_z|+2|J_-|)/2kT}}{Z}$$ } and the concurrence is {\small
$$C_{12}=max \{\frac{\sinh \frac{|J_-|}{kT}-\cosh
\frac{|J_+|}{kT}e^{-|J_z|/kT} }{\cosh \frac{|J_-|}{kT}+ \cosh
\frac{|J_+|}{kT}e^{-|J_z|/kT} },0\}.
$$} Then entanglement occurs when {\small \be f(T)=\sinh \frac{|J_-|}{kT}-\cosh
\frac{|J_+|}{kT}e^{-|J_z|/kT}>0. \label{inequa1}\ee}  This formula
shows that entanglement increase with lowering temperature.
\begin{figure}[h]
\begin{center}
\epsfig{figure=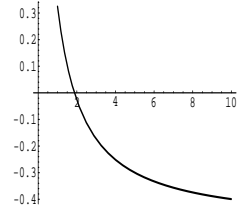,height=3cm,width=3cm}
\end{center}
\caption{Concurrence in $XYZ$ ferromagnet as function of T}
\end{figure}In Fig. 4,
we plot function $f(T)$ for $(J_z,J_y,J_x)=(-3,-2,-1)$. It shows
that entanglement increases with growing anisotropy $J_+$ and
decreases with growing anisotropy $J_-$. Moreover it increases
with growing $J_z$. Thermal entanglement in pure $XYZ$ model has
been studied in \cite{rigolin}, \cite{qiang}. Enhancement of
entanglement in $XYZ$ model in the presence of an external
magnetic field considered in \cite{zhou}, and influence of
intrinsic decoherence on quantum teleportation in \cite{hezhang}.
\subsection{$XYZ$ Model with Magnetic Field ($B = 0 , b=0,
D = 0$)}

The full anisotropic $XYZ$ Heisenberg spin two- qubit system in
which a magnetic field is applied along the $z$-axis, was studied by
Zhou et al. The enhancement of the entanglement for particular fixed
magnetic field by increasing the $z$- component of the coupling
coefficient between the neighboring spins, was their main finding.
\subsection{$XYZ$ Model with DM Coupling ($B = 0 , b=0,
D\neq 0$)} \bea
 \lambda_1&=& \frac{ e^{(-J_z + 2J_-)/2kT}}{Z},\,\,\,
\lambda_2= \frac{ e^{(-J_z - 2J_-)/2kT}}{Z} \\ \lambda_3&=& \frac{
e^{(J_z +2 \nu)/2kT}}{Z},\,\,\,\,\,\,\,\,\,\,\, \lambda_4= \frac{
e^{(J_z -2 \nu)/2kT}}{Z} \eea where $\nu= \sqrt{J_+^2+ D^2}$  \be
Z= 2\left[e^{-J_z/2kT} \cosh \frac{J_-}{kT}+ e^{J_z/2kT} \cosh
\frac{\nu}{kT}\right] \ee

\subsubsection{Antiferromagnetic Case :}
The concurrence is {\small \be C_{12}= max \{ \frac{\sinh \frac
{\nu}{kT}-e^{-J_z/kT} \cosh\frac{J_-}{kT} }{\cosh
\frac{\nu}{kT}+e^{-J_z/kT} \cosh \frac{J_-}{kT}}, 0 \} \ee } and
entanglement occurs when {\small
$$\sinh \frac {\sqrt{J_+^2+ D^2}}{kT}> e^{-J_z/kT}\cosh
\frac{J_-}{kT}.$$} In Fig. 5, we plot the concurrence $C_{12}$ as
function of $D$ and $T$. Comparing with pure $XYZ$ case
(\ref{inequa}), we find that inclusion of DM coupling increases
entanglement.

\begin{figure}[h]
\begin{center}
\epsfig{figure=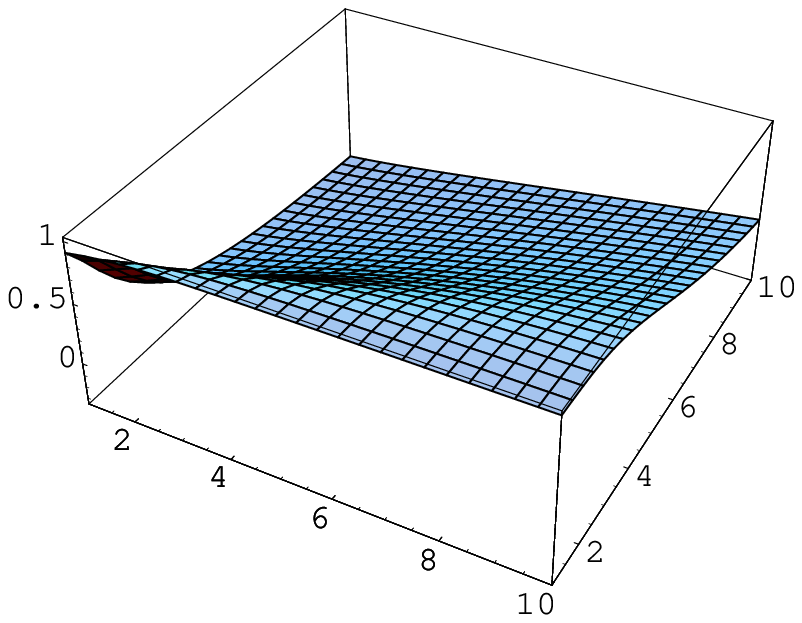,height=3cm,width=3cm}
\end{center}
\caption{Concurrence $C_{12}$ in $XYZ$ antiferromagnet as function
of $D$ and $T$ }
\end{figure}.

\subsubsection{Ferromagnetic Case :} The concurrence is {\small \be
C_{12}= max \{ \frac{\sinh \frac {\nu}{kT}-e^{|J_z|/kT}
\cosh\frac{J_-}{kT} }{\cosh \frac{\nu}{kT}+e^{|J_z|/kT} \cosh
\frac{J_-}{kT}}, 0 \} \ee } and entanglement occurs for
sufficiently strong $D$ {\small
$$\sinh \frac {\sqrt{J_+^2+ D^2}}{kT}> e^{|J_z|/kT}\cosh
\frac{J_-}{kT}.$$} Fig. 6, shows $C_{12}$ as function of $D$ and
$T$.

\begin{figure}[h]
\begin{center}
\epsfig{figure=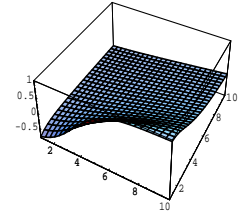,height=3cm,width=3cm}
\end{center}
\caption{Concurrence $C_{12}$ in $XYZ$ ferromagnet as function of
$D$ and $T$ }
\end{figure}

\newpage
\section{Conclusion}
We found in general if $\lambda_1$ (the largest eigenvalue) is
degenerate with $\lambda_2$ then no entanglement occurs. From our
consideration follows that in all cases decreasing of temperature
increases entanglement, if it exists . So that at zero temperature
$T=0$ states are completely entangled $C_{12}=1$. This fact links
entanglement with the Mattis- Lieb \cite{lieb} theorem on absence of
phase transitions in one dimension at $T\neq 0$. Moreover, inclusion
of the DM coupling always increases entanglement, this is why it
could be an efficient control parameter of the entanglement. Our
results show existence of intrinsic relation between weak
ferromagnetism of mainly antiferromagnetic crystals and spin
arrangement in (anti)ferromagnets of low symmetry with entanglement.

Very recently thermal entanglement of a two-qubit isotropic
Heisenberg chain in presence of the Dzyaloshinski-Moriya
anisotropic antisymmetric interaction and entanglement
teleportation, when using two independent Heisenberg $XXX$ chains
as quantum channel, have been investigated \cite{zhang}. It was
found that the DM interaction can excite the entanglement and
teleportation fidelity .  As was noticed $DM$ interaction could be
significant in designing spin-based quantum computers
\cite{kavokin}. Moreover, studying the effect of a phase shift on
amount transferable two-spin entanglement in a spin chain
\cite{maruyama}, it was shown that maximum attainable entanglement
enhanced by $DM$ interaction.

Therefore would be interesting to consider most general $XYZ$
Heisenberg models with DM interaction as quantum channel for
quantum teleportation  which requires to know dependence of
pairwise entanglement on the number of qubits in the spin chain.
These questions now are under investigation.

\begin{acknowledgments}
One of the authors (Z.N.G.) would like to thank Dr. Koji Maruyama
for his helpful remarks. This work was  supported partially by
Izmir Institute of Technology, Turkey.
\end{acknowledgments}

\end{document}